\documentclass[aps,prb,twocolumn,showpacs,superscriptaddress,citeautoscript]{revtex4-1}
\usepackage{amsmath,amssymb,graphicx}
\usepackage{times}

\begin{document}

\title{Finite-size nanowire at a surface: unconventional power laws of
  the van der Waals interaction}

\author {K. A. Makhnovets}
\affiliation{Institute of High Technologies, 
Taras Shevchenko National University of Kyiv,  03022 Kyiv, Ukraine}

\author {A. K. Kolezhuk}
\affiliation{Institute of High Technologies, 
Taras Shevchenko National University of Kyiv,  03022 Kyiv, Ukraine}
\affiliation{Institute of Magnetism, National Academy of Sciences 
and Ministry of Education and Science,  03142 Kyiv, Ukraine}

\begin{abstract}
We study the van der Waals interaction of a metallic or
narrow-gap semiconducting nanowire  with a surface, in the regime of
intermediate wire-surface distances $(v_{F}/c)L \ll d \ll L $ or
$L \ll d \ll (c/v_{F})L $, where $L$ is the nanowire length, $d$ is
the distance to the surface, and
$v_{F}$ is the characteristic velocity of nanowire electrons (for a
metallic wire, it is the Fermi velocity). Our approach, based
on the Luttinger liquid framework, allows one to analyze the
dependence of the interaction on the interplay between the nanowire
length, wire-surface distance, and characteristic length scales
related to the spectral gap and temperature. We show that this
interplay leads to nontrivial modifications of the power law that
governs van der Waals forces, in particular to a non-monotonic
dependence of the power law exponent on the wire-surface separation.
\end{abstract}

\maketitle

\section{Introduction}
\label{sec:intro}

Studies of interactions originating from electromagnetic fluctuations,
the so-called ``dispersion forces'' that include the van der Waals
(vdW) and the Casimir force, have recently experienced a surge due to
their importance for modern material science and technology (see,
e.g.,
Refs.\ \onlinecite{Woods+16rev,Parsegian-book06,Dalvit+book11,Tkatchenko-rev15,Reilly-Tkatchenko-rev15,Ambrosetti+16}
for a review).  Those forces, universally present between any types of
objects, are responsible for the stability of various materials with
chemically inert components. They are especially important at micro-
and nanoscale, playing a significant role in such diverse areas as
catalysis \cite{Norskov+09}, molecular electronics \cite{Moth+09},
nanomechanics, self-assembly \cite{Bartels10,Singh+14}, and biological
phenomena.

vdW forces are generally long-range (decaying as a power of the
distance $d$ between objects), and exponents governing those power
laws are important and convenient characteristics of such
interactions.  For a long time, it was a common practice to treat vdW
forces on the basis of the approximation that describes the coupling
as a sum over pairwise interactions between local fluctuating dipoles.
Although it is well known that vdW interactions are not exactly
pairwise additive, in many cases such an approximation delivers good
results \cite{Parsegian-book06}.
A number of recent studies, however, have revealed several scenarios
with considerable deviations from the ``conventional'' pairwise additive
approximation, involving low-dimensional systems
with reduced or zero spectral gaps (metallic and narrow-gap semiconducting nanowires 
\cite{Chang+71,Glasser72,Dobson+06,Misquitta+10,Misquitta+14,Ambrosetti+16}, carbon nanotubes, graphene
sheets \cite{Dobson+14}). Usually,  only infinite-length
systems are amenable to the analytical
treatment, while for
finite systems and complicated geometries one has to resort to
numerical methods; in particular, the 
many-body dispersion method \cite{Tkatchenko+12} has proved to be 
successful for a wide range of interacting systems \cite{Tkatchenko-rev15,Ambrosetti+16,Woods+16rev}.

In the present paper, we study the vdW interaction in the system of a  metallic or
narrow-gap semiconducting nanowire of a finite size $L$, at distance $d$ to a dielectric or perfect metal
surface, see Fig.\ \ref{fig:wire}. 
To our knowledge, wire-surface interaction has previously been studied
\cite{Emig+06,Bordag+06,Bordag06,Noruzifar+11,Noruzifar+12} only for
the model of an \emph{infinitely long} metallic cylinder interacting
with a metallic plate (half-plane).
The effects of finite length of the wire in
this approach can be studied only as corrections \cite{Bordag06}.  In
contrast to that, we consider the vdW (non-retarded) regime, which, as
we will see below, corresponds to
intermediate wire-surface distances $(v_{F}/c)L \ll d \ll L $ or
$L \ll d \ll (c/v_{F})L $, where
$v_{F}\sim 10^{-2}c$ is the characteristic velocity of nanowire electrons (for a
metallic wire, it is the Fermi velocity); obviously, this regime is
non-existent in the $L\to\infty$ limit.  

In the non-retarded regime, correct description of the nanowire
dynamics is essential.  We are interested in the case of a strongly
one-dimensional (1d) wire, such as a carbon nanotube or a single
polymer molecule. It is well-known that electrons in 1d metallic
systems are not correctly described by the Fermi liquid
\cite{giamarchi-book}, and the proper framework is given by the
Luttinger liquid model.  To describe charge fluctuations in a
finite-size nanowire, we use the Luttinger liquid model with open
boundary conditions, which enables us to study analytically the
behavior of the vdW interaction in different regimes determined by the
interplay between the nanowire length $L$, wire-surface distance $d$,
and characteristic length scales related to the spectral gap and
temperature.  We show that this interplay leads to nontrivial
modifications of the power law that governs the vdW
force. Particularly, we show that at finite temperature the effective
vdW power law exponent can depend on the wire-surface separation in a
non-monotonic way, provided that the spectral gap is sufficiently
small. The paper is organized as follows: Section \ref{sec:model}
outlines the model and the approach utilizing the Luttinger liquid
formalism, in Section \ref{sec:vdW} we analyze the asymptotic behavior
of the vdW potential in various regimes (considering separately the
cases of zero and at finite temperature and spectral gap), and Section
\ref{sec:disc} contains the discussion of our results and a brief
summary.

\section{Model and method}
\label{sec:model}

\begin{figure}[tb]
\includegraphics[width=0.3\textwidth]{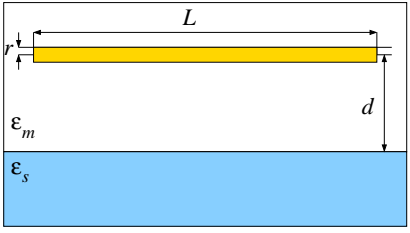}
\caption{
\label{fig:wire}(Color online). Schematic view of the model
  considered: a finite-length nanowire immersed
in the medium with the dielectric constant $\varepsilon_{m}$,
interacting with the substrate with the dielectric constant
$\varepsilon_{s}$.
}
\end{figure}

Consider a nanowire of length $L$ and radius $r\ll L$, placed inside a
medium with the dielectric constant $\varepsilon_{m}$, at distance $d$
to the interface with a substrate with the dielectric constant
$\varepsilon_{s}$, as shown in Fig.\ \ref{fig:wire} (although we
primarily purport to consider insulating substrates, the case of a
perfect metallic substrate can be included by setting
$\varepsilon_{s}\to -\infty$). Assume first that the nanowire is
metallic (discussion of semiconducting nanowires with a small spectral
gap is postponed to Sect.\ \ref{subsec:fingap}).  The low-energy
plasmons in the nanowire can be described in the Luttinger liquid
framework.  Without taking into account the long-range Coulomb
interaction between the electrons, the Hamiltonian of the nanowire can
be written in the language of ``bosonization'' \cite{giamarchi-book}
as follows:
\begin{equation} 
\label{ham-LL} 
\hat{H}_{LL}=\frac{\hbar v_{F}}{2}\int_{0}^{L} dx \big[(\partial_{x}\varphi)^{2}+\Pi^{2} \big],
\end{equation}
where the bosonic field $\varphi(x)$ is related to the charge density
$\rho(x)$ via
\begin{equation} 
\label{rho}
\rho(x)=-e\sqrt{\frac{K}{\pi}}\partial_{x}\varphi, 
\end{equation}
$\Pi(x)$ is the momentum conjugate to $\varphi(x)$, $e$ is the
electron charge, $K$ is the so-called Luttinger parameter which
incorporates the effects of short-range interactions between electrons
($K=1$ for non-interacting electrons, $K<1$ for the case of
short-range repulsion, and $K>1$ for short-range attraction), and
$v_{F}$ is the charge Fermi velocity.  We assume that, as usual, in
the leading order spin and charge degrees of freedom are decoupled
\cite{giamarchi-book}, so spin modes are not included in our
description since we are interested only in charge
fluctuations. Further, for the moment we disregard umklapp processes
that may lead to a gap in the charge sector at a commensurate electron
band filling \cite{giamarchi-book}. If such a gap is smaller than the
"finite-size gap" $\hbar\pi v_F/L$, it may obviously be neglected, and
we will later show that the vdW power law quickly becomes
"conventional" if the gap becomes larger than $\hbar\pi v_F/\min(L,d)$
(see Sect.\ \ref{subsec:fingap} below).

For fixed boundary conditions $\varphi(0)=\varphi(L)=0$,
$\Pi(0)=\Pi(L)=0$ the field operators can be expanded as follows:
\begin{equation} 
\label{fields} 
\Big( \varphi(x), \Pi(x)\Big)=\sqrt{\frac{2}{L}}\sum_{n=1}^{N}\Big(\varphi_{n},\Pi_{n}\Big)\sin\frac{\pi n
  x}{L},
\end{equation}
where it is implied that $n=0$ zero modes are absent in an open system, and the
summation cutoff $N$ is about the number of elementary cells, $N\sim
L/a$ ($a$ is the lattice constant that hereafter will be set to
unity). After this expansion we have
\begin{equation} 
\label{ham-1} 
\hat{H}_{LL}=\frac{\hbar v_{F}}{2}\sum_{n>0}\Big\{
\Pi_{n}^{2}+\Big(\frac{\pi n}{L}\Big)^{2}\varphi_{n}^{2}\Big\},
\end{equation}
and, after introducing the bosonic creation and annihilation
operators  $a_{n}^{\dag}, a_{n}$,   
\begin{equation}
\label{ops}
\varphi_{n}=\sqrt{\frac{L}{2\pi n}}(a_{n}^{\dag}+a_{n}),\quad
\Pi_{n}=i\sqrt{\frac{\pi n}{2L}}(a_{n}^{\dag}-a_{n}),
\end{equation}
 the Luttinger liquid
Hamiltonian (\ref{ham-LL}) takes the familiar form 
\begin{equation} 
\label{ham-0} 
\hat{H}_{LL}
=\frac{\pi \hbar v_{F}}{L} \sum_{n>0}n\Big(a_{n}^{\dag}a_{n} +\frac{1}{2}\Big).
\end{equation}
It should be noted that fixed boundary conditions for $\varphi$ and
$\Pi$ fields used above are not a
standard way of implementing boundaries in a Luttinger liquid
(requiring the electron wave function to vanish at the boundary
results in a more complicated condition involving both $\varphi$ and
$\Pi$); however, in the present work we are not interested in
describing edge states or any other boundary phenomena. Our sole
purpose is to take into account the finite size of the wire, and such
simplified boundary conditions are fully sufficient for that purpose;
previously, such an approach has been successfully used for studying
the conductivity of finite chains \cite{Greschner+13}.

Including the contribution of the long-range Coulomb interactions
between the electrons, we
can write the full
Hamiltonian as $\hat{H}=\hat{H}_{LL}+\hat{U}_{1}+\hat{U}_{2}$, where 
\begin{equation} 
\label{U1} 
\hat{U}_{1}=\frac{\varepsilon_{m}-\varepsilon_{s}}{8\pi
  \epsilon_{0}\varepsilon_{m}(\varepsilon_{m}+\varepsilon_{s})} 
\int_{0}^{L} \!\! dx\int_{0}^{L}\!\! dx' \frac{\rho(x)\rho(x')}{\sqrt{(x-x')^{2}+(2d)^{2}}}
\end{equation}
describes the interaction of electrons with the interface \cite{Jackson-book}, and
\begin{equation} 
\label{U2} 
\hat{U}_{2}=\frac{1}{8\pi
  \epsilon_{0}\varepsilon_{m}} 
\int_{0}^{L} \!\! dx\int_{0}^{L}\!\! dx' \frac{\rho(x)\rho(x')}{\sqrt{(x-x')^{2}+(2r_{0})^{2}}}
\end{equation}
corresponds to the direct Coulomb interaction between the electrons of
the nanowire, regularized at small distances to avoid unphysical
singularity.  Expressions (\ref{U1}), (\ref{U2}) are written using
the electrodynamic Green's function in the non-retarded approximation,
i.e., assuming that the typical wave length of
electromagnetic fluctuations contributing to the interaction energy is
much larger than $\max(L,d)$; we comment on the applicability limits of this approximation later.

The regularization length $r_{0}$ reflects the fact that the electron
wave function in a nanowire is actually delocalized over some length
which is about the wire diameter, and thus the singularity in the
Coulomb interaction gets cured. The precise form of the wave function
depends on the microscopic details of the confining potential, and
thus the "cutoff" is not exactly the wire diameter but a
phenomenological quantity proportional to it.  In what follows, for
the sake of simplicity we set $r_{0}$ to the wire radius $r$; as we
will see later, $r$ enters the results only via log corrections. This
way of regularizing the Coulomb interaction in a nanowire has been
introduced in Ref.\ \onlinecite{GoldGhazali90} and is by now standard \cite{giamarchi-book}.

We use a simplified model describing the
substrate by a single parameter, dielectric constant, implicitly
assumed to be frequency-independent.  Although this model is certainly
insufficient for a quantitative description, one can expect that it
captures the essential physics of the vdW (non-retarded) regime, since the main
contribution to the interaction energy in this regime comes from
the range of relatively low frequencies where the dielectric
   constant of a typical covalent insulator weakly depends on the frequency.

The resulting Hamiltonian including interactions takes the following form:
\begin{eqnarray} 
\label{ham-bosonic}
\hat{H}=\frac{\pi\hbar v_{F}}{L}\sum_{nn'}\Big\{
T_{nn'}\big(a^{\dag}_{n}a_{n'} &+&\frac{\delta_{nn'}}{2}\big)\\
&+&\frac{1}{2}\big(\Phi_{nn'}a_{n}^{\dag}a_{n'}^{\dag} + \text{h.c.}\big)
\Big\},\nonumber
\end{eqnarray} 
where the matrix elements are given by 
\begin{eqnarray} 
\label{TPhi} 
\Phi_{nn'}&=&\frac{1}{2}\sqrt{nn'}\big[b_{2}I_{nn'}(s_{2})-b_{1}I_{nn'}(s_{1}) \big],\nonumber\\
T_{nn'}&=&n\delta_{nn'}+\Phi_{nn'},
\end{eqnarray}
and we have introduced the notation
\begin{eqnarray} 
&&I_{nn'}(s)=\frac{1}{\pi^{2}}\int_{0}^{\pi}dt \int_{0}^{\pi}dt'\;
\frac{\cos(nt)\cos(nt')}{\sqrt{(t-t')^{2}+s^{2}}}, \label{Inn1} \\
&&b_{1}=\frac{\varepsilon_{s}-\varepsilon_{m}}{\varepsilon_{s}+\varepsilon_{m}}b_{2},\quad
b_{2}=\frac{2K\widetilde{\alpha}}{\pi\varepsilon_{m}}, \quad
\widetilde{\alpha}=\frac{e^{2}}{4\pi\epsilon_{0}\hbar v_{F}},\nonumber\\
&& s_{1}=2\pi d/L,\quad s_{2}=2\pi r/L.\label{not}
\end{eqnarray}
Here $\widetilde{\alpha}$ is the analog of the fine structure constant, with the
speed of light $c$ replaced by the Fermi velocity $v_{F}$. Taking into account
that typically $v_{F}\sim 10^{-2}c$, one
concludes that typical values of $b_{1,2}$ are of the order of
unity (or less if $\varepsilon_{m}$ is large). Properties of functions
$I_{nn'}(s)$ are listed in the Appendix.

The quadratic Hamiltonian (\ref{ham-bosonic}) is readily diagonalized
by the Bogoliubov transformation: 
\begin{equation} 
\label{ham-diag} 
\hat{H}=\frac{\pi\hbar v_{F}}{L}\sum_{n>0}
\lambda_{n}\big(A^{\dag}_{n}A_{n}+\frac{1}{2}\big) 
\end{equation}
where $A_{n}$ are the unitary transformed bosonic operators, and
$\lambda_{n}$ are positive solutions of the secular equation
\begin{equation} 
\label{secular} 
\det(M-\lambda^{2}\openone)=0,\quad M=(T+\Phi)(T-\Phi).
\end{equation}
 Although $\lambda_{n}$ can be easily found
numerically, below we demonstrate that a very good approximation is obtained by neglecting non-diagonal
elements of $M$ altogether:
\begin{equation} 
\label{lambda-diag} 
\lambda_{n}\approx \sqrt{M_{nn}}=n\big[1+b_{2}I_{nn}(s_{2})-b_{1}I_{nn}(s_{1}) \big]^{1/2}.
\end{equation}
This is explained by the fact that functions $I_{nn'}(s)$ decay rather
fast with $|n-n'|$ (see the Appendix), so the ratio of non-diagonal to diagonal
elements of $M$ is practically a small parameter, and the leading
contribution to $\lambda_{n}$ from the non-diagonal part of $M$ comes
in the second order in this small parameter.

\section{The vdW interaction potential}
\label{sec:vdW}

Having obtained the spectrum of the interacting Hamiltonian, we are
now in a position to calculate the interaction potential as a function
of the distance $d$ to the surface. At zero temperature, the vdW
interaction energy $W(d)$ is simply the difference between the ground state
energies taken at finite $d$ and at infinity:
\begin{equation} 
\label{vdw-0} 
W_{0}(d)=\frac{\pi\hbar
  v_{F}}{2L}\sum_{n}\big(\lambda_{n}-\lambda_{n}^{(0)} \big),\quad \lambda_{n}^{(0)}\equiv\lambda_{n}(d=\infty).
\end{equation}
At finite temperature $T\not=0$ the vdW energy can be found as the
corresponding difference of free energies, yielding
\begin{eqnarray} 
\label{vdw-T} 
W(d)&=&W_{0}(d)+W_{T}(d),\nonumber\\
W_{T}(d)&=&=T \sum_{n}\ln\left[
\frac{1-e^{-{\ell_{T}\lambda_{n}}/{L} }}
{1-e^{-{\ell_{T}\lambda_{n}^{(0)}}/{L} }}
\right],
\end{eqnarray}
where
\begin{equation} 
\label{elt} 
\ell_{T}=\pi\hbar v_{F}/T
\end{equation}
is the characteristic thermal length.

In the pairwise additive approximation, the vdW interaction would always decay as $W(d)\propto d^{-3}$.
The actual behavior of the interaction energy depends
on the interplay of the four
characteristic problem
scales: the wire length $L$, distance to the surface $d$, the
thermal length $\ell_{T}$, and the characteristic scale
$\ell_{\Delta}$ related to the spectral gap (see Eq.\ \ref{lD} below).
In all our calculations, we will
assume that the distance to the surface is large compared to
the wire radius, $d\gg r$.

\subsection{The vdW interaction at $T=0$}
\label{subsec:zeroT}

\begin{figure}[tb]
\includegraphics[width=0.47\textwidth]{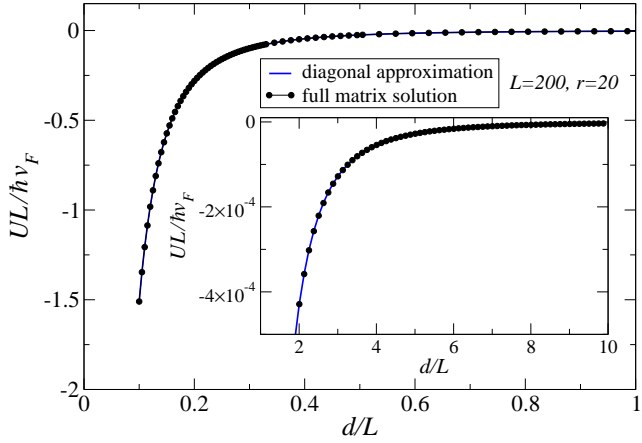}
\caption{\label{fig:diag}(Color online). Typical zero-temperature vdW
  interaction energy calculated via the full diagonalization of the
  matrix $M$ (\ref{secular}) (symbols), compared to the results of
  using the
  diagonal approximation (\ref{lambda-diag}). The parameters taken are $b_{1}=b_{2}=1$.
}
\end{figure}

Consider first the properties of the dispersive interaction energy at
zero temperature. As shown below, those results will be applicable at
finite temperatures as well, as long as $\ell_{T} \gg \min(d,L)$.
In Fig.\ \ref{fig:diag} we compare the results for
the interaction energy
obtained by exact numerical diagonalization of the matrix $M$ (see
Eq.\ (\ref{secular})) with those using the simple diagonal
approximation (\ref{lambda-diag}): one can see that the diagonal
approximation does a very good job, and we have checked this behavior
for various ratios $L/r$. Thus, from now on we will adopt the diagonal approximation,
assuming (\ref{lambda-diag}) for $\lambda_{n}$, and setting
\begin{equation} 
\label{lambda-diag0} 
\lambda_{n}^{(0)}\approx n\big[1+b_{2}I_{nn}(s_{2}) \big]^{1/2}.
\end{equation}

\begin{figure}[tb]
\includegraphics[width=0.47\textwidth]{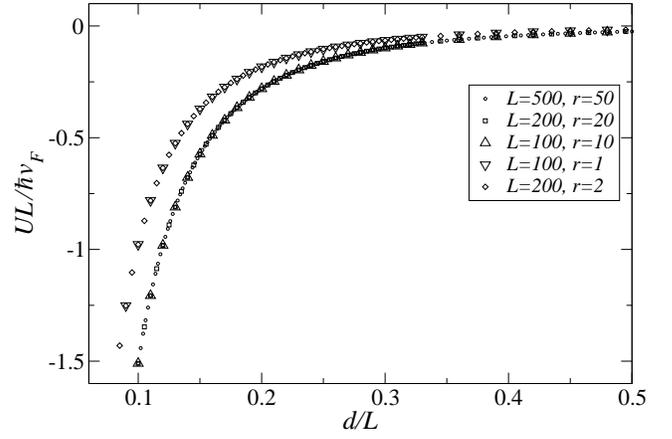}
\caption{\label{fig:Ud} Zero-temperature vdW
  interaction energy at $b_{1}=b_{2}=1$,
expressed in units of
$\hbar v_{F}/L$ as a function of the dimensionless distance $d/L$,
for several values of the nanowire length $L$ and radius $r$, with the
elongation factor $L/r$ is kept fixed.  
}
\end{figure}

Asymptotic behavior of $W_{0}(d)$ can be easily analyzed in a few
limiting cases. Since we assume $r\ll d$ and thus $s_{2}\ll s_{1}$, and $b_{1}$ and
$b_{2}$ are of the same order of magnitude, it follows that 
$b_{1}I_{nn}(s_{1}) \ll b_{2}I_{nn}(s_{2})$. 
Therefore, one may expand (\ref{lambda-diag}) in $b_{1}I_{nn}(s_{1})$,
which yields
\begin{equation} 
\label{w0} 
W_{0}(d)\approx -\frac{\pi \hbar v_{F}}{4L}\sum_{n>0} n\frac{b_{1} I_{nn}(s_{1})}{\sqrt{1+b_{2}I_{nn}(s_{2})}}.
\end{equation}
This expression shows that the vdW energy, expressed in units of
$\hbar v_{F}/L$ as a function of the dimensionless distance $d/L$,
depends only on the geometric factor $L/r$ (at fixed $b_{1,2}$). This scaling is illustrated
in Fig.\ \ref{fig:Ud}.

\subsubsection{$2\pi d\ll L$.}

For $2\pi d\ll L$, the integrals $I_{nn}(s_{1,2})$ in Eq.~(\ref{w0}) can be
replaced by Macdonald's functions (see Eq.\ (\ref{Isll1})), so
the summation is effectively cut off at $n\sim L/d $ (the summand
decays exponentially for larger $n$). One can easily estimate the sum
by passing to an integral; for $b_{2}\ln(d/r) \gg 1$ we obtain
\begin{equation} 
\label{W0d-sll1} 
\frac{W_{0}(d)}{L}\simeq -A \frac{b_{1}}{\sqrt{b_{2}}}
\frac{\hbar v_{F}}{d^{2}\ln^{1/2}(\zeta d/r)},
\end{equation}
where $A$ and $\zeta$ are some numbers of the order of unity.
One can show that taking into account subleading terms in
(\ref{Isll1}) yields merely a correction of the
order of $d/L$  to (\ref{W0d-sll1}). 

It is easy to see from Eq.~(\ref{w0}) that the main contribution to
the interaction for $2\pi d \ll L$
is made by the modes with  $2\pi n d/L \sim 1$, which corresponds to
frequencies $\nu\sim v_{F}/2\pi d$ (e.g., for  $d>10$~nm the relevant
frequency range  is $\nu \sim 10$~THz; in this region the dielectric
   constant of a typical covalent insulator should only weakly depend on the
   frequency). Thus, retardation
effects remain negligible as long as the corresponding typical electromagnetic wave length
$2\pi d(c/v_{F})$ remains large compared to $L$. This determines the
range of distances where the non-retarded vdW regime is realized,
$L(v_{F}/c) \ll 2\pi d \ll L$.

For $b_{2}\ln(d/r)\ll 1$ (since the ``fine structure constant''
$\tilde{\alpha}\sim 1$, this  regime might be
realized only if $\varepsilon_{m}\gg 1$) the term involving $b_{2}$  in 
(\ref{w0}) can be neglected, and the result is
\begin{equation} 
\label{W0d-sll1a} 
\frac{W_{0}(d)}{L}\simeq \widetilde{A}b_{1}\frac{\hbar v_{F}}{d^{2}},
\end{equation}
where $\widetilde{A}\sim 1$ is another numerical factor.

\subsubsection{$2\pi d\gg L$.}

In this case,  $I_{nn}(s_{1})$  in Eq.\ (\ref{w0}) can be replaced by
its asymptotics $4/(\pi n^{4}s_{1}^{3})$, see Eq.\ (\ref{Isgg1}), which
yields the standard $1/d^{3}$ behavior for the vdW energy:
\begin{equation} 
\label{W0d-sgg1} 
\frac{W_{0}(d)}{L}\simeq
-\frac{b_{1}}{8\pi^{3}}
f\Big(b_{2},\frac{r}{L}\Big) 
\frac{L\hbar v_{F}}{d^{3}}.
\end{equation}
Here the function $f(b_{2},r/L)$,  defined as 
\begin{equation} 
\label{f1} 
f\Big(b_{2},\frac{r}{L}\Big) =\sum_{n\geq 1}
\frac{1}{n^{3}}\frac{1}{\sqrt{1+b_{2}K_{0}(2\pi nr/L )}},
\end{equation}
weakly depends on its arguments, being always a number of the
order of unity (see Fig.\ \ref{fig:scaling}).

As seen from the above formula, for $2\pi d\gg L$ the main contribution to the interaction energy comes
from the modes with low frequencies $\nu \lesssim v_{F}/L$, so the
non-retarded approximation remains valid if the typical electromagnetic wave length
$L(c/v_{F})$ is large compared to $d$. Thus, non-retarded regime is in
this case realized in the distance range
$L/(2\pi) \ll d \ll (c/v_{F})L$. 

A comparison of Eqs.\ (\ref{W0d-sgg1}) and (\ref{W0d-sll1}) shows that
the vdW power law changes from $W\sim d^{-3}$ at large distances $d\gg L$ to
$W\sim d^{-2}/\ln(d/r)$  at small distances $d\ll L$. 

\subsection{Finite temperature effects}
\label{subsec:finT}

The summation in
(\ref{vdw-T}) is effectively cut off at $n\sim n_{T}=L/\ell_{T}$, so 
the thermal contribution $W_{T}(d)$ to the vdW energy is exponentially
small if $\ell_{T} \gg L$. 
We therefore assume that the opposite condition of high temperatures $\ell_{T} \ll L$ is
satisfied, which renders the contribution of thermal fluctuations into
the form
\begin{equation} 
\label{wT1}
W_{T}\simeq  -T \sum_{n=1}^{n_{T}}\ln \frac{\lambda_{n}^{(0)}}{\lambda_{n}}.
\end{equation}
Expanding
in $b_{1}I_{nn}(s_{1})$ as done before for $T=0$ case, one
can obtain an estimate for the thermal contribution in the following
form which is easy to analyze:
\begin{equation} 
\label{wT2}
W_{T}(d)\simeq -\frac{T}{2} \sum_{n=1}^{n_{T}} \frac{b_{1}I_{nn}(s_{1})}{1+b_{2}I_{nn}(s_{2})}.
\end{equation}
Below we will see that the
thermal contribution dominates over the quantum one when $\ell_{T}\ll \min(d,L)$. 

\subsubsection{$2\pi d\ll L$.}

For $d \ll \ell_{T} \ll L$, the
thermal contribution is roughly a factor $(d/\ell_{T})^{2}$ smaller
than the quantum one (\ref{W0d-sll1}), and thus can be neglected.
For $\ell_{T}\ll d \ll L$, the estimate yields 
\begin{equation} 
\label{WdT-sll1} 
\frac{W_{T}(d)}{L}\simeq 
\begin{cases}
\displaystyle
-B\left(\frac{\varepsilon_{s}-\varepsilon_{m}}{\varepsilon_{s}+\varepsilon_{m}}\right)
\frac{T}{d\ln(\kappa d/r)}, & b_{2}\ln(d/r) \gg 1 \\
\displaystyle -\widetilde{B} b_{1}
(T/d), & b_{2}\ln(d/r) \ll 1
\end{cases},
\end{equation}
where $B$, $\widetilde{B}$, and $\kappa$ are numerical factors
$~1$. It is easy to see that in this regime the contribution of
thermal fluctuations
is much larger than the corresponding contribution of the ground state
energy (\ref{W0d-sll1}), (\ref{W0d-sll1a}). 
The condition of applicability of non-retarded approximation remains
the same as in the $T=0$ case, namely $L(v_{F}/c) \ll 2\pi d \ll L$.

It is interesting to note that in the retarded regime $L\to\infty$
Emig \emph{et al.} \cite{Emig+06} obtained an expression for the
interaction energy that is similar to the first line of
Eq.\ (\ref{WdT-sll1}) but has a different logarithmic factor
($\ln(c\ell_{T}/v_{F}r)$ instead of $\ln(d/r)$).

\subsubsection{$2\pi d\gg L$.}
\begin{figure}[tb]
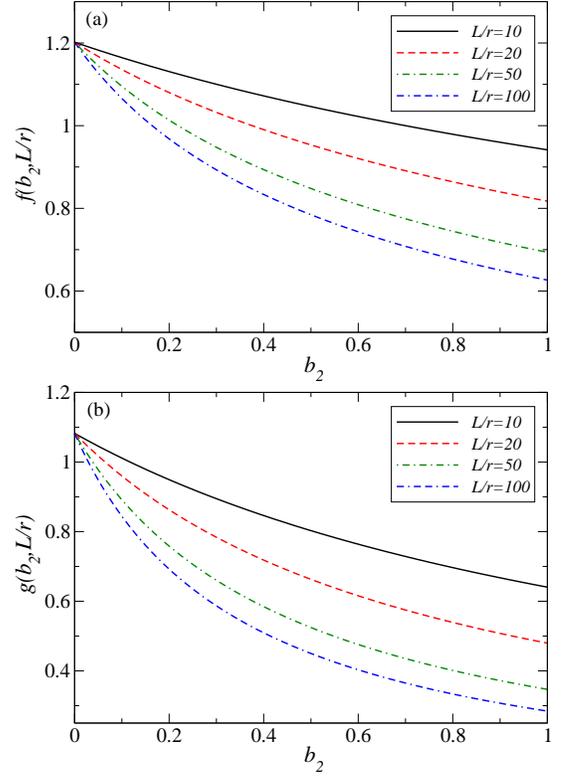

\includegraphics[width=0.4\textwidth]{fbLtor}
\includegraphics[width=0.4\textwidth]{ftbLtor}
\caption{\label{fig:scaling} (Color online).
Plots of scaling functions $f(b_{2})$ and $g(b_{2})$, defined
by Eqs.\ (\ref{f1}) and (\ref{f2}), respectively, for several values
of the elongation factor $L/r$.
}
\end{figure}

If $L \ll \ell_{T} \ll d$, the
thermal contribution is negligible as discussed above, so the only
regime where thermal fluctuations dominate the vdW energy is
$\ell_{T}\ll L\ll d$:
\begin{equation} 
\label{WdT-sgg1}  
\frac{W_{T}(d)}{L}\simeq -\frac{b_{1}}{4\pi^{4}} \frac{TL^{2}}{ d^{3}}
g(b_{2},r/L), \quad  \ell_{T}\ll L\ll d .
\end{equation}
Here function $g(b_{2},r/L)$ is
defined as follows:
\begin{equation} 
\label{f2} 
g\Big(b_{2},\frac{r}{L}\Big) =\sum_{n\geq 1}
\frac{1}{n^{4}}\frac{1}{1+b_{2}K_{0}(2\pi nr/L )},
\end{equation}
and is presented in Fig.\ \ref{fig:scaling} for a few fixed values of $L/r$.
Again, the condition of applicability of non-retarded approximation remains
the same as in the $d\gg L$ case at $T=0$, namely $L/(2\pi) \ll d \ll (c/v_{F})L$.
  
\subsection{Effects of a finite spectral gap}
\label{subsec:fingap}

Although our formalism is based on bosonization and therefore is tuned to
metallic nanowires, it is easy to incorporate effects of a small
spectral gap $\Delta$ and thus generalize our calculations to the case of
narrow-gap semiconducting wires. Indeed, to introduce a gap, one has to add the ``mass
term'' 
\[
\frac{1}{2}\hbar v_{F}\int dx\, (\Delta/\hbar v_{F})^{2} \varphi^{2}
\]
to the Hamiltonian (\ref{ham-LL}). Then, in Eqs.\ 
(\ref{ham-1}), (\ref{ops}), and (\ref{ham-0})  one has to make the
replacement $n\mapsto \sqrt{n^{2}+n_{\Delta}^{2}}$, where
\begin{equation} 
\label{lD}
\ell_{\Delta}=\frac{\pi\hbar v_{F}}{\Delta}, \quad n_{\Delta}=\frac{L}{\ell_{\Delta}},
\end{equation}
$\ell_{\Delta}$ being a new
characteristic length scale related to the gap. The full Hamiltonian
keeps the form (\ref{ham-bosonic}), with the amplitudes modified as follows:
\begin{eqnarray} 
\label{TPhi-gap} 
\Phi_{nn'}&=&\frac{1}{2}\frac{nn' \big[b_{2}I_{nn'}(s_{2})-b_{1}I_{nn'}(s_{1}) \big]}{[(n^{2}+n_{\Delta}^{2})(n'^{2}+n_{\Delta}^{2})]^{1/4}},\nonumber\\
T_{nn'}&=&\sqrt{n^{2}+n_{\Delta}^{2}}\delta_{nn'}+\Phi_{nn'}.
\end{eqnarray}
The diagonal approximation (\ref{lambda-diag}) for $\lambda_{n}$ gets
modified accordingly:
\begin{equation} 
\label{diag-gap} 
\lambda_{n}\simeq \big\{ n_{\Delta}^{2}+n^{2}\big[1+b_{2}I_{nn}(s_{2})-b_{1}I_{nn}(s_{1}) \big] \big\}^{1/2}.
\end{equation}

Our analysis shows (see Fig.\ \ref{fig:power}) that the effect of a finite gap becomes dominating if the condition
$\ell_{\Delta}\ll \min(d,L)$ is satisfied, and in this case the
behavior  of the vdW interaction energy is governed by the standard
power law (as given by the pairwise additive approximation)
$W(d)\propto d^{-3}$.
Thus, introduction of a spectral
gap (i.e., making the wire insulating) leads to the same power law
exponent that corresponds to large  distances $d\gg L$.
It should be noted that this effect is similar to that obtained by
Dobson \textit{et al.} \cite{Dobson+06} for two parallel wires, where
the behavior of the vdW energy changed from $W\sim d^{-5}$ for
insulating wires to $W\sim d^{-2}/\ln^{3/2}(d/r)$ for metallic ones (in
fact, the latter expression has been obtained much earlier by other
authors \cite{Chang+71,Glasser72}). 

\section{Discussion and summary}
\label{sec:disc}

\begin{figure}[tb]
\includegraphics[width=0.4\textwidth]{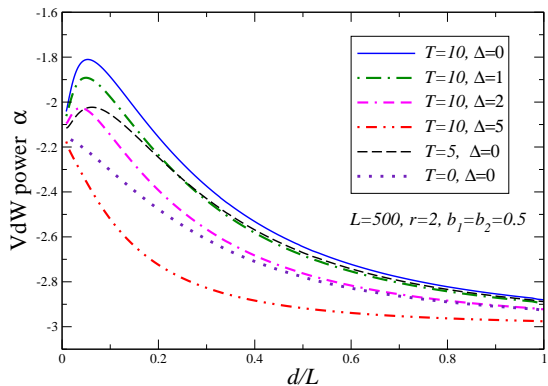}
\caption{\label{fig:power} (Color online). Behavior of the vdW power
  law exponent defined by Eq.~(\ref{alpha}) as a function of the
  wire-surface separation $d$, for different values of the temperature
  $T$ and the spectral gap $\Delta$ expressed in the units of $\hbar
  v_{F}/L$. Non-monotonic behavior of $\alpha$ is suppressed with the
  increase of the gap. 
}
\end{figure}

The analytical estimates obtained in the previous section suggest that
if one describes the vdW interaction between a wire and a surface by
means of a power law $W(d)\propto d^{\alpha}$, then the ``running exponent''
\begin{equation} 
\label{alpha}
 \alpha=\frac{\partial \ln W}{\partial \ln d}
\end{equation}
is a complicated function of the distance
$d$. For instance, at $T=0$ the asymptotics (\ref{W0d-sll1}),
(\ref{W0d-sgg1}) show that this exponent must monotonically change  from $\alpha=-3$ at large
distances $d\gg L$ to $\alpha=-2$ at small distances $d\ll L$. If
temperature is high enough to make $\ell_{T} \ll L$, according to
Eqs.\ (\ref{WdT-sll1}), (\ref{WdT-sgg1}) the behavior of $\alpha$ should be non-monotonic,
changing from $\alpha=-3$ at $d\gg L$, to $\alpha=-1$ at 
$\ell_{T}\ll d \ll L$ and back to $\alpha=-2$ at $d\ll \ell_{T}$.

This is indeed demonstrated in Fig.\ \ref{fig:power}, where we present
the results of numerical calculations using Eqs.\ (\ref{vdw-0}),
(\ref{vdw-T}) in the diagonal approximation for $\lambda_{n}$. One can
see that an increase of the spectral gap leads to a rapid suppression
of  ``unconventional'' power laws, squeezing the corresponding range
of distances.  This is in line with the general idea that
unconventional power laws stem from delocalized excitations
\cite{Dobson+06,Ambrosetti+16}, while the presence of a gap introduces a finite
correlation length.  Further, one can see that a non-monotonic
behavior of $\alpha$ as a function of $d$ is observed only as long as the gap
is sufficiently small, so that
$\ell_{\Delta}\gg \ell_{T}$.

The ``conventional'' $1/d^{3}$ dependence of the vdW interaction
energy can be understood as the result of pairwise summation of the
standard $1/d^6$ potential of vdW interaction between the elements of
the wire and surface viewed as point-like particles.  This result is
valid for a gapped wire if the gap is larger than $\hbar\pi
v_F/\min(L,d)$.  For a wire with smaller gap, as we have shown, this
pairwise additive result is correct only if the wire is far enough
from the surface to be itself approximately considered as a point-like
object (practically if $d > L$).  For shorter distances $d<L$ the
gapless and delocalized character of nanowire excitations becomes
important (the essential point is that the wire susceptibility has
strong dependence both on the frequency and the wave
vector\cite{Dobson+06,Ambrosetti+16}), and as the result the power law
exponent $\alpha$ changes from $-3$ to ``nearly $-2$" or ``nearly
$-1$" (``nearly'' meaning the presence of log corrections),
respectively to whether the energetic or enthropic contribution is
dominating. 
Since the energetic contribution always wins at distances $d$
smaller that the characteristic thermal length $\ell_{T}$, for finite
temperature there are three distinct regimes of $\alpha$ going from
$-3$ to $-1$ and back to $-2$ with the decreasing distance $d$.
 Generally, for $d<L$ there are
logarithmic corrections to the power law which stem from the Coulomb
interaction between electrons inside the nanowire. Those log
corrections may become negligible if the parameter $b_2$ is small
enough; however, since the ``fine structure constant''
$\tilde{\alpha}\sim 1$, this regime might be realized only if the
dielectric constant of the medium is large, $\varepsilon_{m}\gg 1$.

It should be remarked that a non-monotonic behavior of the vdW
power law exponent has been observed in the many-body dispersion (MBD)
calculation \cite{Ambrosetti+16} for the vdW interaction between two
parallel carbyne wires. However, the origin of such a behavior
is different from our case since the calculation in
Ref.\ \onlinecite{Ambrosetti+16} has been performed for $T=0$; in their
results, non-monotonicity reveals itself only at extremely small distances $d/L
\lesssim 10^{-2}$ and is probably connected with details of the
specific realization of the MBD model. In our case, non-monotonicity
is the effect of finite temperature which sets in at distances
$d/L\sim 0.1$.

To summarize, we have applied the Luttinger liquid
approach to study the vdW interaction between a finite-size metallic or
narrow-gap semiconductor nanowire and an insulating or perfect-metal
surface.  
We focused on the case of a strongly
one-dimensional wire, such as a carbon nanotube or a single
polymer molecule, which is not correctly described by the Fermi liquid. 
We obtained simple analytical expressions describing the vdW
interaction in different regimes determined by the interplay between 
characteristic length scales set by the spectral gap and temperature,
the nanowire length, and the wire-surface distance. It is shown that 
the effective vdW power law exponent is generally a complicated function of
wire-surface distance, which can be non-monotonic if the gap is small
enough compared to the temperature. 

It should be emphasized that our results are obtained in the
previously unexplored non-retarded regime that is realized at
intermediate wire-surface distances
\begin{equation}
\label{distances}
(v_{F}/c)L \ll d \ll L \quad  \text{or} \quad L \ll d \ll (c/v_{F})L ,
\end{equation}
where
$v_{F}\sim 10^{-2}c$ is the characteristic velocity of nanowire electrons (for a
metallic wire, it is the Fermi velocity).
For that reason, they cannot be directly related to the results of
previous studies
\cite{Emig+06,Bordag+06,Bordag06,Noruzifar+11,Noruzifar+12} since
those were obtained for the model of an infinitely long wire
($L\to\infty$).  It is interesting to note that in the $d\ll L$ case
our $T=0$ result (\ref{W0d-sll1}) for the interaction energy has the
same functional dependence on the distance $d$ as the retarded-regime
expression obtained by Noruzifar \emph{et al.}
\cite{Noruzifar+11,Noruzifar+12} for a "plasma cylinder" interacting
with a perfect metal plate; however, those two results correspond to
different physics, as is obvious from the fact that the prefactor in
the result of Noruzifar \emph{et al.}  vanishes when the cylinder
radius goes to zero, while in our case the prefactor contains the
Fermi velocity and does not depend on the wire radius. 

There are further similarities between our results and those obtained
previously in the retarded regime. For example, in the ``universal''
limit $d/r\to\infty$, $L=\infty$ the dispersive interaction energy
is \cite{Emig+06,Noruzifar+11} $W=-\hbar c L/[16\pi d^{2}\ln(d/r)]$,
which is ``nearly $\alpha=-2$'' power law as our Eq.\ (\ref{W0d-sll1}), but contains $c$ instead
of $v_{F}$ and has a different power in the logarithmic correction. 
In the high-temperature retarded regime Emig \emph{et al.} \cite{Emig+06} obtained an expression for the
interaction energy which is similar to the first line of
Eq.\ (\ref{WdT-sll1})  but again has a different logarithmic factor,
$\ln(c\ell_{T}/v_{F}r)$ instead of $\ln(d/r)$. Although the
underlying physics, as we emphasized before, is different, those similarities in functional
dependence of the interaction energy on distance
stem from the analogous mathematical structure of summing over gapless 
modes with linear quasi-1d dispersion: in our case those modes are plasmons, in
the retarded regime they are electromagnetic waves. The origin of log
corrections in the retarded regime is different as well: they arise due to
the logarithmic behavior of the 1d propagator in the limit of low wave
vectors.

Finally, we would like to remark on the origin of similarities between
our results, obtained by the direct microscopic description of the wire in the
bosonization framework, and those of  Dobson \emph{et
  al.} \cite{Dobson+06} who studied interaction between two parallel infinite-length
wires using RPA expressions for the wire response: (i) in one dimension,
as it is well known \cite{DzyaloshinskiiLarkin73}, RPA  gives essentially exact results for the
density-density correlation function, which are the same  as in the
Luttinger liquid approach \cite{giamarchi-book}, and (ii) obviously the
interaction between two wires is sufficiently similar to the interaction
between a wire and its ``mirror image'' below the surface (which is
what the non-retarded approximation essentially reduces to).  We have 
checked that results of Ref.\ \onlinecite{Dobson+06} are reproduced in
the approach of two interacting Luttinger liquids (with the additional
bonus of the ability to consider finite-length wires), but this is
outside the scope of the present work.


\acknowledgments
We thank V. Lozovski for helpful discussions. This work has been supported
by the grant 16BF07-02 from the Ministry of Education and Science of Ukraine.

\appendix
\section*{Appendix: Properties of $I_{nn'}(s)$}

Here we list some properties of functions $I_{nn'}(s)$
defined in (\ref{Inn1}). Passing from $y$, $y'$ to new variables $y\pm
y'$, one can perform one integration and rewrite the integral as
\begin{widetext}
\begin{equation} 
\label{I1}
I_{nn'}(s)= \begin{cases}
0, & |n-n'|=\text{odd} \\  \displaystyle
-\frac{2}{\pi (n^{2}-n'^{2})}\int_{0}^{\pi/s} dz
\frac{n\sin(nsz)-n'\sin(n'sz)}{\sqrt{1+z^{2}}}, & n\not=n'\\
 \displaystyle \frac{1}{\pi
  n}\int_{0}^{\pi/s}dz\frac{n(\pi-sz)\cos(nsz)-\sin(nsz)}{\sqrt{1+z^{2}}},
& n=n'.
\end{cases}
\end{equation}
\end{widetext}
The asymptotic expressions for small and large argument are easily
obtained. For $s\gg 1$ the leading asymptotics are
\begin{equation} 
\label{Isgg1} 
I_{nn'}(s)\approx\begin{cases}
\displaystyle\frac{6\pi}{n^{2}n'^{2}s^{5}} +O(s^{-7}), & \text{$n$ and $n'$ are even} \\ 
\displaystyle \frac{4}{\pi n^{2}n'^{2}s^{3}} +O(s^{-5}), & \text{$n$ and $n'$ are odd}
\end{cases},
\end{equation} 
and for $s\ll 1$ one has
\begin{equation} 
\label{Isll1} 
I_{nn'}(s)\approx\begin{cases}\displaystyle -\frac{1}{n+n'}+\frac{2s}{\pi} +O(s^{2}), &
\text{$n\not= n'$} \\
\displaystyle K_{0}(ns)-\frac{\text{Si}(\pi n)}{\pi n} +O(s^{2}), & \text{$n=n'$,}
\end{cases}
\end{equation}
where $K_{0}(z)$ is the modified Bessel function of the second kind
(the Macdonald function), and $\text{Si}(z)$ is the sine integral function.



\begin{thebibliography}{99}
\bibitem{note1} In this case, there are always regions of the wire
and of the surface separated by a sufficiently large
distance $R$ such that the characteristic propagation time $R/c$ is
larger than the typical period of local charge density
oscillations, so retardation should be taken into account.
\end{thebibliography}
\end{document}